\documentclass[twocolumn,showpacs,aps,amsmath,amssymb]{revtex4}
\usepackage{bm,color}%\usepackage{graphicx}

\newcommand{\beq}{\begin{equation}}
\newcommand{\eeq}{\end{equation}}
\newcommand{\bqa}{\begin{eqnarray}}
\newcommand{\eqa}{\end{eqnarray}}

\definecolor{ngreen}{rgb}{0.2,0.6,0.2}

\definecolor{golden}{rgb}{0.8,0.6,0.1}

\begin{document}
\title{Comment on ``Quantum phase for an arbitrary system with\\ finite-dimensional Hilbert space''} 
\author{Michael J. W. Hall and David T. Pegg}
\affiliation{Centre for Quantum Computation and Communication Techology (Australian Research Council), Centre for Quantum Dynamics, Griffith University, Brisbane, QLD 4111, Australia}
%\date{\today}

% USE FOR BOTH
\begin{abstract}
A construction of covariant quantum phase observables, for Hamiltonians with a finite number of energy eigenvalues, has been recently given by D. Arsenovi\'{c} {\it et al.} [Phys. Rev. A {\bf 85}, 044103 (2012)].  For Hamiltonians generating periodic evolution, we show that this construction is just a simple rescaling of the known canonical `time' or `age' observable, with the period $T$ rescaled to $2\pi$.  Further, for  Hamiltonians generating quasiperiodic evolution, we note that the construction leads to a phase observable having several undesirable features, including (i) having a trivially uniform probability density for any state of the system, (ii) not reducing to the periodic case in an appropriate limit, and (iii) not having any clear generalisation to an infinite energy spectrum.  In contrast, we note that a covariant time observable has been previously defined for such Hamiltonians, which avoids these features.  We also show how this `quasiperiodic' time observable can be represented as the well-defined limit of a sequence of periodic time observables.
\end{abstract}

%USE FOR REVTEX
\maketitle

\section{Introduction}

Physical systems that change in time, whether they are classical or quantum, can potentially be used as clocks.  In particular, they have observable quantities which are correlated with, and hence provide information about, the passage of time.  Such quantities may be referred to as `time observables' or, to better distinguish them from the fundamental time parameter used in evolution equations, as `age observables' \cite{pegg98}.

For quantum systems, the construction of canonical time observables is well known, both for Hamiltonians generating periodic evolution \cite{pegg98, disc,discon}, and for Hamiltonians with continuous energy spectra \cite{discon,cont}.  These observables are characterised by having optimal resolution properties, under any energy constraint, for covariantly tracking the passage of time \cite{hall}.  Here, covariance refers to the property that the corresponding `time' probability density, $p(t|\psi)$, is simply translated under evolution of the system \cite{disc,discon}, i.e., 
\begin{equation} \label{cov}
   p(t|\psi_\tau) = p(t-\tau|\psi_0),
\end{equation}
where $\psi_\tau$ denotes the state of the system at evolution time $\tau$.  Thus, if $p(t|\psi_0)$ is initially peaked about $t=0$,  then $p(t|\psi_\tau)$ is peaked about $t=\tau$. 

If the Hamiltonian has a discrete spectrum with incommensurate energy differences, then the evolution of the system will generally not be periodic, although it will return arbitrarily closely to its initial state an infinite number of times \cite{nonper}. Such systems are said to be almost periodic, or quasiperiodic.  The canonical covariant time observable for a quasiperiodic system has been given recently \cite{hall}, generalising the periodic case.  The construction applies to any discrete energy spectrum, whether finite or infinite, and whether degenerate or nondegenerate \cite{hall}.

Surprisingly, in light of the above, Arsenovi\'{c} {\it et al.} have recently stated that  ``a definition of the phase observable for an arbitrary quantum system with a periodic or quasiperiodic state vector dynamics has not been formulated in full generality,'' where `phase' refers to a variable  ``which is directly related to the time parameter'' \cite{ars}.  They then give  constructions of covariant phase observables for particular cases.

In Sec.~II below we show that, for periodic systems, the construction of Arsenovi\'{c} {\it et al.} is  just a simple rescaling of the known canonical time observable, with the period $T$  rescaled to $2\pi$.  

Further, for quasiperiodic systems, we note in Sec.~III that that the construction of Arsenovi\'{c} {\it et al.} yields an observable with several undesirable features, including having a trivial uniform probability density for any state of the system.  This feature implies that the observable does not yield any information whatsoever about the evolution time parameter, and is essentially due to  restricting the domain of the observable to a finite interval -- which we show is inappropriate both for classical and quantum quasiperiodic systems.  We also briefly indicate how the canonical time observable defined in Ref.~\cite{hall} avoids such undesirable features.  

Finally, we demonstrate in Sec.~IV that the canonical time observable of a quasiperiodic system may be represented as the well defined limit of a sequence of periodic time observables.

\section{Periodic systems}

For the case of periodic evolution, under a Hamiltonian $H$ with nondegenerate energy eigenstates $\{|E_0\rangle,|E_1\rangle,\dots\}$, the probability density of the corresponding canonical time observable is given for state $|\psi\rangle=\sum_n c_n|E_n\rangle$ by \cite{pegg98}
\begin{equation} \label{per}
  p_T(t|\psi) = \frac{1}{T} \left| \sum_n c_n e^{iE_nt/\hbar} \right|^2 ,
  \end{equation}
where $T$ denotes the period of the evolution.  This probability density is periodic, and  normalised over any reference interval of length $T$, such as $[0,T)$.  It is well known for the particular case of a one-dimensional harmonic oscillator \cite{disc,discon}, and appears to have first been explicitly given for the general case in Ref.~\cite{pegg98} (where it is called the `age' observable). 

The corresponding positive operator valued measure (POVM), $\{A_t\}$, for the canonical time observable, or age, 
%of a nondegenerate periodic system,  
follows immediately from Eq.~(\ref{per}) as \cite{pegg98}
\begin{equation} \label{perpovm} 
A_t = \frac{1}{T} \sum_{m,n} |E_m\rangle\langle E_n|\, e^{-i(E_m-E_n)t/\hbar} . 
\end{equation}
Integration of $A_t$ over the interval from $t_a$ to $t_b$ yields the 
 semispectral measure $M_t^{p_k/q_k}(t_a,t_b)$ given in the first line of Eq.~(5) of Ref.~\cite{ars} (noting that $T=2\pi\hbar/\Delta E_k$ in the notation of the latter).  Moreover, the phase observable for periodic systems defined in Eq.~(5) of Ref.~\cite{ars} is just a rescaling of this measure, by a factor $2\pi/T$, mapping the time interval $[0,T)$ to the phase interval $[0,2\pi)$.  Thus, the periodic phase observable defined by Arsenovi\'{c} {\it et al.} is simply a trivial rescaling of the known canonical time observable for periodic systems. This rescaling does, however, have some advantage in giving an immediate comparison of the fractions of a cycle completed by systems of different frequencies.

\section{Quasiperiodic systems}
 
The phase semispectral measure in Ref.~\cite{ars} represents the probability of an outcome in the interval $[t_a, t_b]$, which is a finite fraction of the period T, whereas the age observable in Eq. (3) refers to an infinitesimal interval.  This leads to difficulties for the former in the quasiperiodic case, as the semispectral measure then represents the probability of an outcome in an infinite time interval, which washes out all fine structure. 

Essentially for this reason, the phase observable defined by Arsenovi\'{c} {\it et al.} for the quasiperiodic case has several unsatisfactory properties, discussed below.  We also show how the corresponding canonical time observable  avoids these difficulties.

\subsection{Three problems}

First, the statistics of the phase observable defined by Arsenovi\'{c} {\it et al.} are purely random, corresponding to a uniform distribution over the interval $[0,2\pi)$ for any state of the system (see Eqs. (8) and (9) of \cite{ars}). This is due to the washing out of fine structure as noted above.  It follows in particular that the phase observable contains {\it no} information about {\it any} properties of the system, including any evolution properties.

The underlying reason is that Arsenovi\'{c} {\it et al.} restrict attention, {\it a priori}, to probability densities defined on the finite interval $[0,2\pi)$.  However, such a restriction is inappropriate for quasiperiodic systems -- even in the classical case.  As a simple example, consider a quasiperiodic classical system with action-angle variables $(\phi_1,\phi_2,J_1,J_2)$ \cite{arnold}, and time dependence
\[ \phi_k(t)=\omega_k \,t\!\! \mod 2\pi,~~~~J_k(t)=J_k(0),~~~~k=1,2 \]
such that the frequencies $\omega_1$ and $\omega_2$ are incommensurate, i.e., $\omega_1/\omega_2$ is irrational.  It immediately follows that there is a one-one mapping between the angles $(\phi_1,\phi_2)$ and the evolution time $t$ \cite{oneone}, i.e., that there is a classical `time observable' of the form
    \[ t = t_C = f(\phi_1,\phi_2). \]
Hence, despite the angles $\phi_1$ and $\phi_2$ each being restricted to values in $[0,2\pi)$, the corresponding time observable is uniquely determined as a function of these phase space  observables, and takes values in $(-\infty,\infty)$.  

It follows that, even classically, any observable that tracks the time evolution of a quasiperiodic system is expected to have an infinite range of possible values.  However, ``the defining property of the phase observable'' in equation (1) of Ref.~\cite{ars} (equivalent to the time tracking property of covariance in Eq.~(\ref{cov}) above),  {\it a priori} restricts the evolution parameter $\theta$, appearing in the time evolution operator $e^{-iH\theta}$, to a finite interval. This is clearly inappropriate in light of the above, and no reason for this restriction is given by Arsenovi\'{c} {\it et al.}  In contrast,  the canonical time observable for a periodic quantum system has an infinite range of values \cite{hall}, as expected by analogy with the classical case.  As will be seen below, this leads to  nontrivial statistics in general, with a uniform distribution only in the case of an energy eigenstate.

Second, the statistics of the quasiperiodic phase observable defined by Arsenovi\'{c} {\it et al.} are  uniform even for those states of the system which evolve {\it periodically}.   As an example, for the three-level quantum system considered in Sec.~II of Ref.~\cite{ars}, in the quasiperiodic case where $\nu:=(E_2-E_1)/(E_1-E_0)$ is irrational, suppose that the initial state has the form
\begin{equation} \label{ex} 
|\psi_0\rangle = c_0|E_0\rangle + c_1|E_1\rangle . 
\end{equation}
The evolution of this state is periodic, with period $T=2\pi\hbar/(E_1-E_0)$.  Moreover, this periodic evolution is invariant under any perturbation of the Hamiltonian by $\epsilon|E_2\rangle\langle E_2|$. It follows that (i) for any  value of $\epsilon$ such that $\nu(\epsilon):=(E_2+\epsilon-E_1)/(E_1-E_0)$ is rational, Eqs.~(6) and (7) of Ref.~\cite{ars} uniquely define a  {\it nonuniform}  phase distribution for this state, while (ii) for any value of $\epsilon$ such that $\nu(\epsilon)$ is irrational, Eqs.~(8) and (9) of Ref.~\cite{ars} uniquely define a {\it uniform} phase distribution for  this state. Yet the evolution is identical in both cases!   

It is clearly undesirable, and arguably physically inconsistent, that the `phase' properties of a given state, $|\psi_0\rangle$, are not determined by its evolution {\it per se}, but in a discontinuous manner according to the arbitrary choice of a parameter under which the state and its evolution are invariant.  
In contrast, as will be seen below, the statistics of the canonical time observable only depend on the state and its evolution, and reduce to periodic statistics for the above example.  

Third, Arsenovi\'{c} {\it et al.} state in Sec.~III of Ref.~\cite{ars} that ``more discussion is needed to treat systems
with degenerate energy and/or irrational ratios of the energy eigenvalue differences,'' and also that ``systems with an infinite Hilbert space and an infinite number of discrete energy eigenvalues such that the energy spectrum contains accumulation points requires a careful analysis.''  In contrast, the canonical time observable is well defined in all such cases \cite{hall}. 

\subsection{One solution}

The canonical time observable  for quasiperiodic systems avoids the problems noted above for the quasiperiodic phase observable of Arsenovi\'{c} {\it et al.}  For brevity, only the case of a nondegenerate energy spectrum will be considered here. The canonical quasiperiodic probability density for state $|\psi\rangle =\sum_n c_n|E_n\rangle$ is then defined by \cite{hall}
\begin{equation} \label{qper}
p(t|\psi) := \left| \sum_{n} c_n e^{iE_nt/\hbar}\right|^2 .%= 1 + \sum_{j\neq k} c_j^*c_k e^{-i(E_j-E_k)t/\hbar} .
\end{equation}
While this is very similar to the periodic density defined in Eq.~(\ref{per}), there is a crucial difference: the expectation value of a function $f(t)$ is evaluated via the `almost periodic' measure on the real numbers, $\mu_{ap}$, rather than via the usual Lebesgue measure, with \cite{hall, bohr}
\begin{equation} \label{bes}
\langle f(t)\rangle_\psi =\mu_{ap}[fp] := \lim_{\tau\rightarrow\infty} \frac{1}{\tau} \int_0^{\tau}dt\,f(t)\,\,p(t|\psi) .
\end{equation}
This measure is well defined on the class of quasiperiodic functions \cite{bohr}, i.e., for any function $f(t)$ having a countable Fourier series $f(t)=\sum_k f_ke^{i\omega_kt}$.  Thus, the expectation value of any quasiperiodic function of the canonical time observable can be calculated, analogous to the calculation of the expectation value of any periodic function of a periodic time observable via Eq.~(\ref{per}).

The corresponding quasiperiodic POVM, $\{M_t\}$, corresponding to the canonical time observable, follows from Eq.~(\ref{qper}) as \cite{hall}
\begin{equation} \label{povm}
M_t = \sum_{m,n} |E_m\rangle\langle E_n|\, e^{-i(E_m-E_n)t/\hbar} .
\end{equation}
It is normalised relative to the almost periodic measure, with $\mu_{ap}[M_t]=\hat{1}$, where $\hat{1}$ denotes the unit operator.  Like the canonical time observable for periodic and continuous quantum systems, the quasiperiodic canonical time observable is covariant, and has optimal time resolution properties under any energy constraint \cite{hall}.

To see how the canonical time observable overcomes the problems noted above for the phase observable defined by Arsenovi\'{c} {\it et al.}, note first that the quasiperiodic probability density in Eq.~(\ref{qper}) is defined over the whole interval $(-\infty,\infty)$, and is normalised with respect to the almost periodic meaure $\mu_{ab}$.  This density is uniform if and only if the system is in an energy eigenstate, implying that measurement of the canonical time observable extracts `time' information whenever the system is not stationary, just as one would expect by analogy with the classical case.  In contrast, the quasiperiodic phase observable of Arsenovi\'{c} {\it et al.} always has a uniform probability distribution.
%, normalised via the Lebesgue measure, due to an unjustified and inappropriate restriction of their observable to a finite range.

Further, for the example of Eq.~(\ref{ex}) above, one may use Eqs~(\ref{qper}) and (\ref{bes}) to calculate the expectation value of any function $f(t)$ having period $T=2\pi\hbar/(E_1-E_0)$, yielding

~\\

\begin{eqnarray} \nonumber
 \langle f(t)\rangle_\psi &=& \lim_{N\rightarrow\infty} \frac{1}{NT+\tau'} \int_0^{NT+\tau'} dt\,f(t)\,p(t|\psi) \\ \nonumber
 &=& \lim_{N\rightarrow\infty} \frac{1}{NT} \sum_{j=0}^{N-1} \int_0^T dt\,f(t) \,p(t|\psi)\\ \nonumber
 &=& \frac{1}{T}\int_0^T dt\,f(t) \,p(t|\psi) = \int_0^T dt\, f(t)\, p_T(t|\psi),% \left| \sum_{j=0,1} c_j e^{iE_jt/\hbar}\right|^2 ,
 \end{eqnarray} 
where $\tau'$ is any value in $[0,T)$ and $p_T(t|\psi)$ is defined in Eq.~(\ref{per}).  Thus, it is  identical to the expectation value calculated from the corresponding periodic time observable via Eq.~(\ref{per}).  Hence, in contrast to Ref.~\cite{ars}, the quasiperiodic time observable reduces to the periodic time observable in the appropriate limiting case.

Finally, again in contrast to Ref.~\cite{ars}, the quasiperiodic time observable in Eqs.~(\ref{qper})-(\ref{povm}) generalises straighforwardly to the case of a degenerate spectrum, and is well defined whether or not the energy spectrum is finite or infinite, has commensurate or incommensurate eigenvalue differences, or has an accumulation point \cite{hall}.    It may further be remarked that the particular cases of incommensurate eigenvalue differences (such as an anisotropic oscillator), and an accumulation point (such as a bound hydrogen atom), have quite interesting resolution and information-theoretic properties \cite{hall}.

For completeness, it may also be noted here that a `Hermitian time operator' $T_G$, satisfying the commutation relation $[T_G,H]=i\hbar$  on a dense set of states for a Hamiltonian $H$ with a strictly infinite energy spectrum, has been proposed by Galapon \cite{gal}.  However, this operator does not satisfy the fundamental covariance property (\ref{cov}).  Further, the commutation relation only holds, for any evolving state $|\psi_\tau\rangle$, at a set of times of total measure zero \cite{hallcom}.  Hence this operator does not have any clear interpretational connection to a time observable.  

\section{Quasiperiodic time observables as a limit of periodic time observables}

As seen above, the statistics of the canonical time observable for quasiperiodic systems reduces to those for periodic systems in the appropriate limit.  It is of interest to note that a converse relation may also be obtained, with the quasiperiodic time observable given by a suitable limit of periodic time observables.  This is, necessarily, very different to the limit used by Arsenovi\'{c} {\it et al.} to obtain a quasiperiodic phase observable as a limit of periodic phase observables \cite{ars}.

First, it may be remarked that one can always approximate the evolution of a quasiperiodic system  to some specified accuracy, over any finite interval $[0,\tau)$, by some periodic system having period $T\geq\tau$ \cite{bohr}.  Choosing $\tau$ to be larger than the relevant period of experimental interest (which will be no longer than the age of the universe, and typically rather shorter), then this approximation will be sufficient for all practical purposes.  Hence, in a practical sense, periodic time observables may be considered sufficient.  
Even so, it remains of fundamental interest to determine the limit as the approximation becomes arbitrarily accurate.  By doing so one can in fact dispense with the need for approximation in the first place.

In particular, for a given energy spectrum $\{E_0,E_1,\dots\}$ having incommensurate eigenvalue differences, suppose that one has a sequence of approximations $\{E^{(k)}_0,E^{(k)}_1,\dots\}$  corresponding to periodic evolution of period $T_k$, with $E^{(k)}_n\rightarrow E_n$ and $ T_k\rightarrow \infty$ as $k\rightarrow\infty$.  Further, for any quasiperiodic function $f(t)$ define the corresponding periodic function $f_k(t):= f( t\!\!\mod T_k)$.  If $\langle f_k(t)\rangle_\psi$ denotes the expectation of $f_k(t)$ for the periodic time observable corresponding to period $T_k$, it immediately follows from Eq.~(\ref{per}) that 
\begin{eqnarray} \nonumber
\lim_{k\rightarrow\infty} \langle f_k(t)\rangle_\psi &=& \lim_{k\rightarrow\infty} \frac{1}{T_k}\int_0^{T_k} dt\,f(t)\,\left| \sum_n c_n e^{iE^{(k)}_nt/\hbar} \right|^2\\ \nonumber
&=& \lim_{T\rightarrow\infty} \frac{1}{T} \int_0^T dt\,f(t)\,\left| \sum_n c_n e^{iE_nt/\hbar} \right|^2\\
&=& \langle f(t)\rangle_\psi ,
\end{eqnarray}
where $\langle f(t)\rangle_\psi$ is the expectation value of $f(t)$ for the quasiperiodic time observable in Eq.~(\ref{bes}).  Hence, the latter observable corresponds to the limit of a sequence of periodic time observables, as claimed.

\section{Conclusion}

It has been shown that the periodic phase observable defined by Arsenovi\'{c} {\it et al.} in Ref.~\cite{ars} is a simple rescaling of the known periodic canonical time observable, or age \cite{pegg98}. Further, the quasiperiodic phase observable defined by Arsenovi\'{c} {\it et al.} has several undesirable features, which are avoided by the known quasiperiodic canonical time observable \cite{hall}.  The connections between the periodic and quasiperiodic cases have also been discussed. Finally we should remark that rescaling the age to a phase for periodic systems, as done by Arsenovi\'{c} {\it et al.}, does have some advantage in giving an immediate comparison of the fractions of a cycle completed by systems of different frequencies.

\acknowledgments

This work was supported by the ARC Centre of Excellence CE110001027.


\begin{thebibliography}{99}
\bibitem{oneone} If the angles $\phi_1$ and $\phi_2$ were the same at two different times $t$ and $t'$, one would have $\omega_k(t-t')=2\pi n_k$ for some integers $n_1$ and $n_2$, implying that $\omega_1/\omega_2=n_1/n_2$ is rational, contrary to assumption.
\end{thebibliography}
\end{document}